\documentclass[
 reprint,
 amsmath,amssymb,
 aps,prb,
 showpacs
]{revtex4-1}

\usepackage{natbib}
\usepackage{amsmath}
\usepackage{graphicx}
\usepackage{dcolumn}
\usepackage{epstopdf}
\usepackage{bm}
\usepackage{slashed}

\newcommand{\g}{\gamma}

\begin{document}

\DeclareGraphicsExtensions{.jpg,.JPG,.jpeg,.pdf,.png,.eps}
\graphicspath{{./graphics/}}

\title{Holographic interaction effects on transport in Dirac semimetals}

\author{V.P.J. Jacobs}
 \email{V.P.J.Jacobs@uu.nl}
\author{S.J.G. Vandoren}
\author{H.T.C. Stoof}
\affiliation{
Institute for Theoretical Physics and Center for Extreme Matter and Emergent Phenomena,
Utrecht University, Leuvenlaan 4, 3584 CE Utrecht, The Netherlands
}

\date{\today}

\begin{abstract}
 Strongly interacting Dirac semimetals are investigated using a holographic model especially geared to compute the single-particle correlation function for this case, including both interaction effects and nonzero temperature. We calculate the (homogeneous) electrical conductivity at zero chemical potential, and show that it consists of two contributions. The interband contribution scales as a power law either in frequency or in temperature for low frequency. The precise power is related to a critical exponent of the dual holographic theory, which is a parameter in the model. On top of that we find for nonzero temperatures a Drude peak corresponding to intraband transitions. A behavior similar to Coulomb interactions is recovered as a special limiting case.
 \end{abstract}

\maketitle

\section{Introduction} Dirac semimetals are a state of matter that can be seen as the three-dimensional version of graphene. Indeed, Dirac semimetals are zero-gap semiconductors, and without interactions their conduction and valence band touch each other at isolated points in momentum space, the so-called Dirac points \cite{Herring37}. As a result these semimetals have a low-energy description in terms of massless (3+1)-dimensional Dirac fermions with a linear dispersion. Such a Dirac fermion consists of two Weyl fermions of opposite chirality. Breaking time-reversal or spatial inversion symmetry, the associated degeneracy of the Dirac point is lifted and the two Weyl nodes become separated in momentum-space, forming a Weyl semimetal \cite{Savrasov11,Balents11,Balents12}. This hypothetical phase of matter exhibits unusual transport properties such as an anomalous Hall effect \cite{Ran11} and also gapless surface states forming a Fermi arc instead of the usual closed Fermi surface.

Dirac semimetals have been predicted theoretically \cite{Young12} and are known to occur on the phase boundary between a topological and a trivial insulator \cite{Murakami07}. However, to realize the Dirac semimetal in this manner turned out to be experimentally challenging \cite{Hasan11}. Nevertheless, recent theoretical \cite{Fang12,Wang13} and experimental progress has ultimately resulted in the realization of Dirac semimetals in the crystals Na$_3$Bi \cite{Chen14} and Cd$_3$As$_2$ \cite{Neupane13,Cava13,Chen14v2}. Their crystal symmetry prevents the Dirac points from becoming gapped, making these systems a more robust testing ground for relativistic physics in a tabletop experiment.
On the theoretical side, most of the work on Dirac and Weyl semimetals has been on noninteracting systems \cite{Savrasov11,Balents11,Ran11,Balents12,Tewari13}, although in a number of cases, also Coulomb interactions \cite{Burkov11,Hosur12,Sekine13} and short-ranged interactions \cite{Mastropietro13} have been considered. In this work, however, we focus on a more strongly interacting Dirac semimetal that is coupled to a critical order-parameter field near a quantum critical point. If this critical point is nontrivial, the order-parameter fluctuations may induce strong interactions between the Dirac fermions that are not necessarily of a Coulombic nature and whose treatment goes beyond perturbation theory.
In particular, we want to address what the effective low-energy theory is for Dirac fermions in the presence of such generic critical order-parameter fluctuations. Since the behavior of fermions coupled to a critical collective mode is a long-standing problem in the context of non-Fermi liquids \cite{Saarloos02,Wolfle07}, this motivates the search for a description of the strongly interacting Dirac semimetal using the recently developed techniques from the so-called anti-de Sitter/conformal field theory (AdS/CFT) correspondence \citep{[{For an introduction to applications of the AdS/CFT-correspondence to condensed matter, see e.g. the following lecture notes, and references therein: }]Herzog09,*Hartnoll09,*McGreevy10}. In this work, we indeed present such a holographic model for a Dirac semimetal, where the interactions between the Dirac fermions are mediated by the critical fluctuations modeled in holography by a strongly coupled conformal field theory. The model describes a class of gapless and particle-hole symmetric systems, that behave as Dirac semimetals with strong interactions in the infra-red and which are free in the ultra-violet.

To achieve this, we generalize previous work \cite{ARPES12,Weyl13} to formulate a holographic model that allows us to obtain the single-particle Green's function of the strongly interacting Dirac semimetal. Most importantly for our purposes, this correlation function satisfies the desired (zeroth-order) frequency sum rule, which makes it a feasible candidate for applications in realistic solid-state materials, e.g., by a direct comparison to angle-resolved photoemission spectroscopy (ARPES) experiments. Next, using this single-particle Green's function, we determine also the electrical conductivity including the effects of the holographic interactions and nonzero temperature. It is very important to realize that because of particle-hole symmetry, the fermionic contribution to the electrical conductivity remains finite even in the absence of disorder. This is a result of the fact that in the particle-hole symmetric case the electric field cannot affect the center-of-mass motion of the system and that the interactions cause a drag between the electrons and holes that lead to a finite relaxation time for the charge current.

\begin{figure}[t]
\vskip 10pt
\centering
\includegraphics[width=79mm]{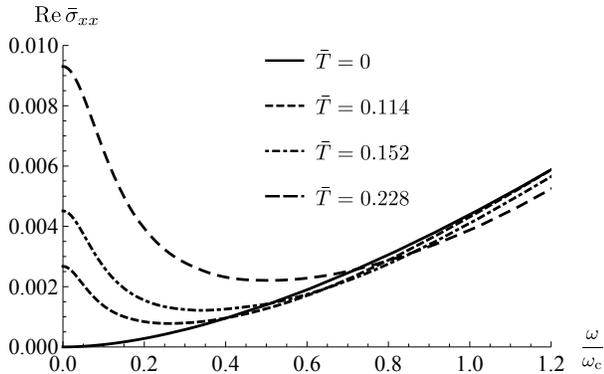}
\caption{Real part of the dimensionless fermionic conductivity $\text{Re}\,\bar{\sigma}_{xx}=\text{Re}\,\sigma_{xx} \hbar c/e^2\omega_{\text{c}}$ as a function of the rescaled frequency $\omega/\omega_{\text{c}}$ defined in the text. The curves are for the holographic parameter $M=1/4$ and for different values of the dimensionless temperature $\bar{T}=k_B T/\hbar \omega_{\text{c}}$. For $\omega/\omega_{\text{c}}\gg 1$, the Dirac semimetal is free and $\sigma \propto |\omega|$ (not visible here). For $T=0$ (solid curve) and as $\omega \simeq \omega_{\text{c}}$, a crossover occurs to a regime where self-energy effects are dominant. Here, the conductivity scales as $\sigma \propto |\omega|^{3-4M}$. For lower frequencies and $T>0$, on top of this power law, a Drude peak appears, corresponding to intraband transitions. For $T>0$ and in the far infra-red, i.e., $\hbar\omega\ll k_B T \ll \hbar \omega_{\text{c}}$, the interactions are temperature dominated and the dc conductivity scales as $\sigma \propto T^{3-4M}$.}
\label{fig:sigma}
\end{figure}

Our main result, the optical conductivity, is plotted in Fig.~{\ref{fig:sigma}}. It consists of two contributions, which can be understood as coming from interband and intraband transitions. In particular, for very high frequencies, we obtain the free result \cite{Savrasov11,Hosur12} where the conductivity scales linearly with frequency. As the frequency is lowered, a cross-over takes place as the holographic interaction effects dominate the conductivity. An example of other work in which there is a cross-over behavior from infra-red to ultra-violet in the context of holography is Ref.~\onlinecite{Gubser09}. For zero temperature and in the infra-red, the interband conductivity scales as $|\omega|^{3-4M}$, where $-1/2<M<1/2$ is the (dimensionless) fermion mass in the Anti-de-Sitter background and physically represents a parameter related to the anomalous dimension of the order-parameter fluctuations in the conformal field theory. For $k_B T\gg \hbar \omega$, the interaction effects are temperature dominated and the interband conductivity scales as $T^{3-4M}$. On top of this, we have a Drude-like peak coming from the intraband contribution.

Interestingly, in the case $M=1/2$, which requires a separate computation, the self-energy scales linearly with logarithmic corrections.\cite{Weyl13} These two features are also present in the conductivity, resulting in a Coulomb-like behavior. Indeed, a linear scaling with logarithmic corrections is precisely the behavior found in Ref.~\onlinecite{Hosur12} for the dc conductivity in the case of Coulomb interactions. Note that the Fermi velocity is equal to the speed of light in our relativistic model so it is not renormalized.

\section{Holography} The AdS/CFT correspondence generally provides correlation functions of operators in a strongly coupled conformal field theory, which can be computed by solving classical equations of motion in the dual curved space-time of one dimension higher \cite{Herzog09,*Hartnoll09,*McGreevy10}.
 In particular, in the spirit of the semiholographic approach \cite{Faulkner11}, the holographic prescription that starts from a free Dirac fermion in a curved (4+1)-dimensional bulk space-time was shown in Refs.~\onlinecite{ARPES12, Weyl13} to lead on the flat (3+1)-dimensional boundary of the bulk space-time to a model that corresponds to an elementary Weyl fermion coupled to a conformal field theory. This is in agreement with the general observation that holographic techniques are particularly appealing for the description of chiral boundary fermions \cite{Iqbal09}. To generalize this prescription to a Dirac fermion, we essentially need two copies of this model, with the two Weyl fermions having opposite chirality. This can be illustrated by the free Dirac Hamiltonian, of which a particular representation in the Weyl basis is $H = \tau_3\otimes \boldsymbol\sigma \cdot c\hbar\mathbf{k}$ with $c$ the speed of light. The last factor is the Weyl Hamiltonian for a single Weyl cone, where $\hbar \boldsymbol\sigma/2$ denotes the electron spin, and the Pauli matrix $\tau_3$ introduces the second Weyl cone with opposite chirality.
We start by considering a (4+1)-dimensional asymptotically Anti-de Sitter background with radius $\ell$, with the line element
\begin{equation}\label{eq:bg}
ds^2= - \frac{V^2(r) r^2}{\ell^2} c^2 dt^2 + \frac{\ell^2}{r^2 V^2(r)}dr^2 +
\frac{r^2}{\ell^2} d\vec{x}^2.
\end{equation}
Here, $r$ is the extra spatial holographic dimension and the system is in thermal equilibrium at a temperature $T=1/k_B\beta$ due to the presence of the planar black hole in the bulk described by the blackening function $V(r)=\sqrt{1-(\pi\ell^2 /\hbar \beta c r)^4}$.
We now consider two uncoupled species of free probe Dirac fermions in 4+1 dimensions with masses $ M_i \hbar/c\ell$, where $i=1,2$ and $M_i$ are dimensionless numbers. They are described by the spinor fields $\Psi^{(i)}$ that propagate in the curved background of Eq.~(\ref{eq:bg}). The Dirac masses are subject to the restriction  $-1/2 \le M_i \le 1/2$ where the cases $M_i = \pm 1/2$ need to be treated separately \cite{Weyl13}. With respect to the boundary chirality operator, both bulk Dirac fields can be conveniently expressed in terms of two chiral spinors $\Psi_{R(L)}^{(i)}$. Using the Dirac equation, the chiral components of each species are expressed in terms of the other as
\begin{equation}\label{eq:dirac}
\Psi^{(i)}_{L} = \left(\begin{array}{cc} 0 & 0 \\ - i \xi^{(i)} & 0\end{array}\right) \Psi^{(i)}_R,
\end{equation}
which is the defining equation for the diagonal 2$\times$2 ~matrices $\xi^{(i)}$. Choosing the masses such that $M_1= -M_2 \equiv M$, the complex eigenvalues of $\xi^{(1)}$ and $\xi^{(2)}$ are $\xi_+$ and $\xi_-$, and $-1/\xi_-$ and $-1/\xi_+$, respectively. A different choice for the masses $M_i$ would not lead to this inverse relationship between the eigenvalues and ultimately break the desired spatial inversion symmetry on the boundary.
As in Ref.~\onlinecite{Iqbal09}, the bulk equations of motion for $\xi_{\pm}$ are given by
\begin{align}\label{eq:xi}
\left(\frac{r}{\ell}\right)^2 V(r) \partial_r \xi_{\pm} &+ \frac{2M r}{\ell^2} \xi_{\pm}  \\ &= \frac{\omega}{c V(r)} \mp |\mathbf{k}| +\left(\frac{\omega}{c V(r)}\pm |\mathbf{k}|\right)\xi_{\pm}^2,\nonumber
\end{align}
with the in-falling boundary condition at the horizon $\xi_{\pm}(\pi \ell^2/ \hbar\beta c) = i$.  The variation of the action leading to Eq.~(\ref{eq:xi}) is well-defined only if we specify the action at the boundary. Therefore, Dirichlet conditions  \cite{Iqbal09} are imposed on half of the chiral components of each bulk Dirac field. Most conveniently for our purposes we fix $\Psi^{(1)}_R$ and $\Psi^{(2)}_L$ at a slice $r=r_0$ close to the boundary, specifying the boundary terms as
\begin{equation*}
S_{\partial}=i g_f\int_{r=r_0} d^4 x \sqrt{-g}\sqrt{g^{rr}}\left(\bar{\Psi}^{(1)}_R \Psi^{(1)}_L-\bar{\Psi}^{(2)}_L\Psi^{(2)}_R\right),
\end{equation*}
where $g$ and $g_{\mu \nu}$ are respectively the determinant and the $\mu\nu$-component of the metric corresponding to Eq.~(\ref{eq:bg}) and $g_f$ is a dimensionless normalization constant. Besides this, following the procedure in Refs. \onlinecite{ARPES12,Weyl13,Faulkner11}, we add kinetic terms for both $\Psi^{(1)}_R$ and $\Psi^{(2)}_L$ on the same slice which do not obstruct the variational principle. These additional terms describe elementary fermionic excitations in the boundary theory,
\begin{equation*}
S_{\text{kin}}= Z \int_{r=r_0} d^4 x \sqrt{-g} \left(\bar{\Psi}^{(1)}_R\slashed{\mathcal{D}} \Psi^{(1)}_R+\bar{\Psi}^{(2)}_L\slashed{\mathcal{D}} \Psi^{(2)}_L\right).
\end{equation*}
Here, $Z$ is a dimensionful constant, $\slashed{\mathcal{D}} = \Gamma^{a} e_{{a}}^{\;\;\mu} i
 \partial_{\mu}$, where $e_{{a}}^{\;\;\mu}$ are the vielbeins corresponding to the metric in Eq.~(\ref{eq:bg}), and $\Gamma^{{a}}$ are the Dirac matrices in the (4+1)-~dimensional bulk.
The holographic prescription instructs that the generating functional for correlation functions in the boundary field theory is equal to the limit $r\rightarrow \infty$ of the on-shell bulk action.
Fourier transforming and putting the action on-shell
using Eq.~(\ref{eq:dirac}), the bulk Dirac action vanishes, but the boundary terms do not. Performing the Gaussian integration over
half of the chiral components of each species boils down to eliminating $\Psi^{(1)}_L$ and $\Psi^{(2)}_R$ using Eq.~(\ref{eq:dirac}). After this, we carry out a field rescaling so that both $\Psi^{(1)}_R$ and $\Psi^{(2)}_L$ acquire the canonical dimensions of a (3+1)-dimensional spinor, and subsequently take a specific double-scaling limit to bring $r_0 \rightarrow\infty$, namely \cite{ARPES12,Weyl13},
\begin{equation*}
\frac{r_0}{\ell}\rightarrow \infty,\qquad g_f \rightarrow 0,\qquad \frac{g_f}{Z} \left(\frac{r_0}{\ell}\right)^{2-2M} \rightarrow \frac{\lambda}{ c\ell^{2M}}.
\end{equation*}
This results in an effective boundary action from which the corresponding retarded Green's function for the (3+1)-dimensional boundary Dirac fermion $\Psi=\Psi^{(1)}_R+\Psi^{(2)}_L$ can be obtained.
So each of the chiral components of this Dirac
fermion is supplied by one of the bulk fermion species. In four-vector notation, using the Minkowski metric with signature $(-1,1,1,1)$, the retarded single-particle correlation function is given by
\begin{equation}\label{eq:greens}
G_R(k) = \frac{c k_{\mu} + \Sigma_{\mu}(k)}{\big( c k + \Sigma(k)\big)^2}\gamma^{\mu}\gamma^0,
\end{equation}
where $c k^0= \omega+i 0\equiv \omega^+$, $\gamma^{\mu}$ are the (boundary) Dirac matrices, and
$\Sigma_{\mu}(k)$ are the components of the effective self-energy of the strongly interacting Dirac semimetal obeying
\begin{align}\label{eq:Sigma}
&\Sigma_{\mu}(k)    \nonumber \\
&=-\frac{\lambda}{2} \lim_{r_0\rightarrow \infty} \left(\frac{r_0}{\ell^2}\right)^{2M}\left[\left(\xi_++\xi_-\right)\delta^0_{\mu} + \left(\xi_+-\xi_-\right)\frac{k_i}{|\vec{k}|}\delta^i_{\mu}\right].
\end{align}
Here, the index $i$ runs over the spatial directions, and $\lambda~\geq~0$ is the square of the coupling constant between the Dirac fermion and the dual conformal field theory containing critical order-parameter fluctuations with an anomalous dimension related to $M$.
Notice that for zero temperature, $V(r)~=~1$, and Eq.~(\ref{eq:xi}) can be solved analytically \cite{Weyl13}, resulting in $\Sigma_{\mu}(k,T=0)= \left(ck/\omega_{\text{c}}\right)^{2M-1} c k_{\mu}$, where $\omega_{\text{c}} =  \left[\lambda \Gamma(1/2-M)/(2c)^{2M}\Gamma(1/2+M)\right]^{1/(1-2M)}$.
Finally, the spectral-weight function is the 4$\times$4 matrix $A(\mathbf{k},\omega) = \textrm{Im}[G_R(k)]/\pi$. After diagonalization, its components are given by
\begin{align}\label{eq:specpm}
&\mathcal{A}_{\pm}(\mathbf{k},\omega)\nonumber \\
&=\frac{1}{\pi} \textrm{Im}\left[\frac{-1}{\omega^+ \mp c |\mathbf{k}| + \lambda\lim_{r_0\rightarrow\infty}(r_0/\ell^2)^{2M}\xi_{\pm}}\right],
\end{align}
where the $+(-)$ component denotes the conduction (valence) band of the Dirac semimetal. The components of the spectral-weight function $\mathcal{A}_{\pm}$ are normalized such that $\int_{-\infty}^{\infty} \mathcal{A}_{\pm}(\mathbf{k},\omega) d\omega = 1$, so the desired frequency sum rule is obeyed. This is in contrast to earlier holographic computations in the literature which yield correlation functions of composite operators \cite{Iqbal09,Zaanen09,Vegh11}.

\section{Electrical conductivity}
The fermionic contribution to the electrical conductivity $\sigma^{\mu\nu}$ is computed in linear-response theory. The (3+1)-dimensional Dirac fermions are thus minimally coupled to a sufficiently small background electric field. The Kubo formula relates the conductivity to the retarded current-current correlation function $\Pi^{\mu\nu}(\mathbf{q},\omega^+)$ in the Dirac semimetal as $\sigma^{\mu
\nu}(\mathbf{q},\omega)~=~i \Pi^{\mu\nu}(\mathbf{q},\omega^+)/\omega$.
The retarded current-current correlation function can be expressed in terms of the fermionic Green's function from Eq.~(\ref{eq:greens}) using the particle-hole bubble diagram. To compute it, we start off in the Matsubara formalism, and afterwards make an analytic continuation to real frequency to obtain the retarded correlation function.
Several cases can now be distinguished.

In the noninteracting case, i.e., $\lambda=0$, the Matsubara Green's function is given by $G_M(k)= k_{\mu} \g^{\mu}\g^0/c k^2$ with $k^0 = i\omega_m/c$ and $\omega_m$ the fermionic Matsubara frequencies. After regularization and analytic continuation, the free current-current correlation function at zero temperature is a manifestly transversal tensor, i.e., $\Pi_0^{\mu\nu}(q) = (q^2\eta^{\mu\nu}-q^{\mu}q^{\nu})\Pi_0(q^2)$. Considering the homogeneous response, we obtain then at zero temperature $\sigma_{0,xx}(\mathbf{0},\omega)= i e^2 |\omega| \log[-(\omega^+/\omega_{\text{exp}})^2] /12\pi^2\hbar c$. The real part of the free conductivity is universal and coincides with the result known from the literature \cite{Savrasov11,Hosur12}, Re[$\sigma_{0,xx}(\mathbf{0},\omega)] = e^2 |\omega|/12\pi\hbar c$, i.e., the free conductivity of two coincident Weyl cones of opposite chirality at zero temperature and Fermi velocity $c$.
The imaginary part of the conductivity, however, is nonuniversal and depends on a single frequency parameter $\omega_{\text{exp}}$ that should be determined by experiment from the vanishing of the imaginary part of the conductivity at that particular frequency in the Dirac semimetal of interest.

The zero temperature result just discussed originates from particle-hole excitations, i.e., transitions between the valence and conduction band. Additionally, at nonzero temperature, the noninteracting conductivity contains a Drude peak of weight $T^2$, which comes from the transport of thermally excited particles and holes within the same band. These two contributions are referred to as the inter- and intraband contribution, respectively. At nonzero temperature, the real part of the total noninteracting conductivity can be computed analytically, the result is
 \begin{align}\label{eq:freecond}&\text{Re}\,\sigma_{0,xx}(\mathbf{0},\omega) \nonumber \\
 &=\frac{e^2}{3\hbar c} \left(\frac{\pi}{3} \left[\frac{k_B T}{\hbar}\right]^2 \delta(\omega) + \frac{|\omega|}{4\pi} \tanh\left[\frac{\hbar |\omega|}{4 k_B T}\right]\right).
\end{align}

In the interacting case, computing the current-current correlation function from the bubble diagram with dressed propagators and including the vertex corrections is an exact approach within linear-response theory. The interacting Green's function is given by Eq.~(\ref{eq:greens}). As an approximation we ignore the vertex corrections, but do take into account the self-energy corrections to the propagator. This approximation is justified and consistent in the particle-hole symmetric case at zero chemical potential that is of interest to us here. This is because in this specific case the conductivity is finite even without impurity scattering, and the vertex corrections do not lead to a qualitatively different behavior of the conductivity. In this so-called $GG$ approximation, the current-current correlation tensor is not manifestly transversal anymore, but it is in fact almost transversal at zero temperature, with a 5\% error. In the $GG$ approximation, the conductivity consists again of two contributions, an interband and intraband part.
Due to rotational invariance, all three spatial components of the conductivity tensor are equal. The total contribution to the interacting conductivity is in the $GG$ approximation given by
\begin{align}
\text{Re}\,\sigma_{xx}(\mathbf{0},\omega) = \sigma^{\text{inter}}(\mathbf{0},\omega)+\sigma^{\text{intra}}(\mathbf{0},\omega)\nonumber.
\end{align}
We discuss the two contributions separately, which makes it easier to quantify the distinct behavior they lead to. The results are also plotted separately in Figs.~\ref{fig:inter} and \ref{fig:intra}.

\subsection{Interband contribution}
\begin{figure}[t]
\centering
\vskip 10pt
\includegraphics[width=83mm]{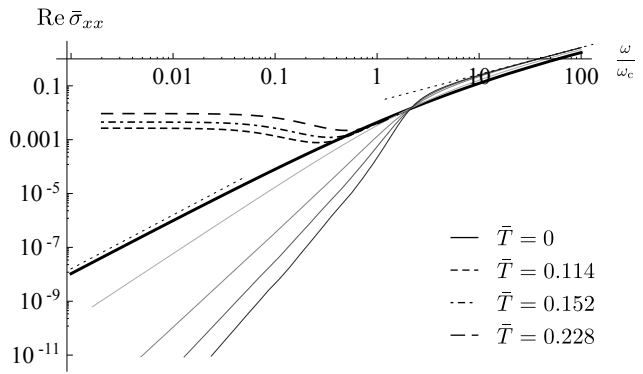}
\caption{Logarithmic plot of the real part of the dimensionless fermionic conductivity $\text{Re}\,\bar{\sigma}_{xx}$ as a function of the rescaled frequency $\omega/\omega_{\text{c}}$ defined in the text. The black curves are for $M=1/4$ and for different values of the dimensionless temperature $\bar{T}$. The grayscale curves are for $T=0$ and for various values of $M$, respectively from top to bottom $M=1/8$, $M=-1/8$, $M=-1/4$ and $M=-1/3$. The dotted lines denote the asymptotics for $M=1/4$, respectively the free result for $\omega/\omega_{\text{c}}\gg1$, and the zero-temperature infra-red result for $\omega/\omega_{\text{c}}\ll 1$. Both results are mentioned explicitly in the text. }
\label{fig:logsigma}
\end{figure}

\begin{figure}[t]
 \centering
\includegraphics[width=78mm]{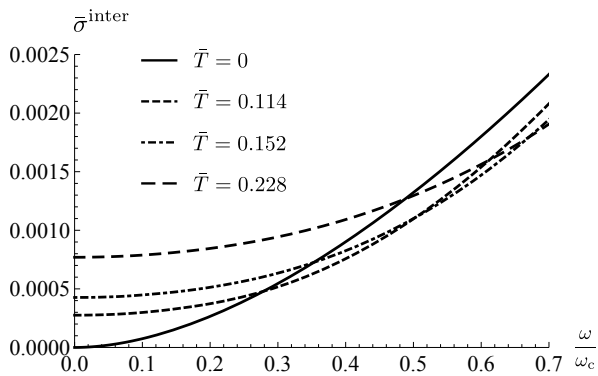}
\caption{The dimensionless purely interband contribution $\bar{\sigma}^{\text{inter}}=\sigma^{\text{inter}} \hbar c/e^2\omega_{\text{c}}$ as a function of the rescaled frequency $\omega/\omega_{\text{c}}$ for $M=1/4$ and for different values of the dimensionless temperature $\bar{T}$. For $T>0$, this contribution goes to a constant as $\omega \rightarrow 0$.}
\label{fig:inter}
\end{figure}

The interband contribution to the interacting conductivity is in the $GG$ approximation given by
\begin{align}
& \sigma^{\text{inter}}(\mathbf{0},\omega)\nonumber \\
& = \frac{2 e^2c^2 }{3\pi}\int_0^{\infty} d\rho \, \rho^2 \int_{-\infty}^{\infty} d\omega' \frac{N_f(\hbar\omega')-N_f(\hbar\omega'+\hbar\omega)}{\hbar\omega}\nonumber \\
& ~\times\Big(\mathcal{A}_+(\rho,\omega') \mathcal{A}_-(\rho,\omega'+\omega)+\mathcal{A}_-(\rho,\omega') \mathcal{A}_+(\rho,\omega'+\omega)  \Big), \nonumber
\end{align}
with $N_f(\epsilon)=(1+e^{\beta \epsilon})^{-1}$. It is shown in Fig.~\ref{fig:inter}.

For zero temperature, two regimes can be discerned, which are reflected in the logarithmic plot of Fig.~\ref{fig:logsigma} as curves with a different slope.\footnote{In Fig.~\ref{fig:logsigma} the total conductivity is shown, but at zero temperature it is dominated by the interband contribution for all frequencies.} For large external frequency, the interband conductivity approaches the ultra-violet behavior, which is the interband part of the free result, i.e., the second term in Eq.~(\ref{eq:freecond}). Here $\sigma$ scales linearly in $\omega$. As the external frequency is decreased, a cross-over occurs to the infra-red behavior. Here, the interband conductivity is dominated by the fermionic self-energy and scales as a different power law in frequency. Indeed, from dimensional arguments, we can infer that the zero-temperature interband conductivity in the infra-red vanishes as $\sigma(\omega) \propto e^2 \omega_{\text{c}}(|\omega|/\omega_{\text{c}})^{3-4M}/12\pi\hbar c $.
This scaling behavior is confirmed by the numerical results shown in Fig.~\ref{fig:logsigma}. The cross-over to the infra-red behavior occurs as the self-energy term in the Green's function becomes dominant over the kinetic term, which is precisely at $\omega_{\text{c}}$.
This low-frequency behavior reflects the fact that the order-parameter field has an anomalous dimension related to $M$ in the infra-red.
An alternative way to understand the behavior of the interband contribution can be accomplished using Fermi's golden rule. The effect of interactions is to smear out the delta peaks in the spectral function. At zero temperature, Lorentz invariance ensures that the delta peaks are broadened only inside the light-cone. As a consequence, an on average slightly higher photon energy is needed to acquire the same scattering rate as in the noninteracting case, so effectively the conductivity is lowered.
For nonzero temperatures, the interband conductivity does not vanish but instead goes to a constant as $\omega\rightarrow0$, as shown in Fig.~\ref{fig:inter}. A nonzero temperature breaks Lorentz invariance and has an additional smearing-out effect on the peaks in the spectral-weight function, this time also outside of the light cone. Now there is residual spectral weight at the Dirac point, so even at zero photon frequency interband transitions can be made, resulting in a nonvanishing interband contribution to the dc conductivity. This effect dominates in the far infra-red where $\hbar \omega \ll k_B T$. Depending on the value of temperature compared to the cross-over frequency $\omega_{\text{c}}$, there are two or three regimes upon increasing the photon frequency. For $k_B T<\hbar \omega_{\text{c}}$, the zero-temperature interaction effects start to dominate, and the interband conductivity approaches the zero-temperature result from below, compensating for the extra smearing out and the spectral weight now available at zero frequency. This explains the intersection with the zero temperature curve in Figs.~\ref{fig:sigma} and \ref{fig:inter}. As the frequency is increased beyond $\omega_{\text{c}}$, the curve will approach the free result as explained above. If $k_B T>\hbar \omega_{\text{c}}$ a cross-over immediately to the free ultra-violet regime takes place around $k_B T =\hbar \omega$, and no intermediate regime can be discerned.
\subsection{Intraband contribution}
\begin{figure}[t]
\centering
\includegraphics[width=78mm]{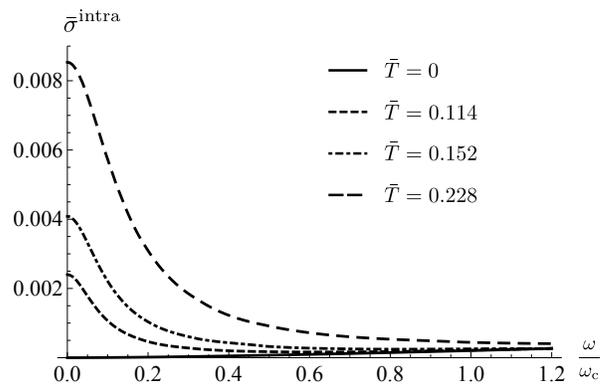}
\caption{The dimensionless purely intraband contribution $\bar{\sigma}^{\text{intra}}=\sigma^{\text{intra}} \hbar c/e^2\omega_{\text{c}}$ as a function of the rescaled frequency $\omega/\omega_{\text{c}}$ for $M=1/4$ and for different values of the dimensionless temperature $\bar{T}$. For nonzero temperature the Drude peak is visible.}
\label{fig:intra}
\end{figure}

The intraband contribution is in the $GG$ approximation given by
\begin{align}
& \sigma^{\text{intra}}(\mathbf{0},\omega) \\
& = \frac{ e^2c^2 }{3\pi}\int_0^{\infty} d\rho \, \rho^2 \int_{-\infty}^{\infty} d\omega' \frac{N_f(\hbar\omega')-N_f(\hbar\omega'+\hbar\omega)}{\hbar\omega}\nonumber \\
& ~\times\Big(\mathcal{A}_+(\rho,\omega') \mathcal{A}_+(\rho,\omega'+\omega)+\mathcal{A}_-(\rho,\omega') \mathcal{A}_-(\rho,\omega'+\omega)  \Big). \nonumber
\end{align}
This contribution is shown in Fig.~\ref{fig:intra}.

In the noninteracting case, only zero frequency transitions contribute to $\sigma^{\text{intra}}$, leading to the first term in Eq.~(\ref{eq:freecond}). If there are interactions, the bands become smeared out and less well-defined. In particular, some spectral weight is moved to the location of the other band. Therefore, even at zero temperature, there is a small contribution from high frequency transitions in $\sigma^{\text{intra}}$, and this leads to a similar power-law behavior in the IR as in the interband part. Namely, $\sigma^{\text{intra}}$ scales as $\omega^{3-4M}$ for small frequency. In the UV, the interactions have a perturbative effect, so that the intraband contribution scales as $\omega^{2M}$. Because this power is less than one, it is subdominant to the interband part for high frequency. Indeed, the total conductivity asymptotes to the noninteracting case where it scales linearly with frequency, as can be seen in Fig.~\ref{fig:logsigma}. The zero temperature intraband contribution vanishes at zero frequency just like the interband contribution.

At nonzero temperature, a Drude peak appears in the IR on top of the power law behavior just discussed. It is the interaction analog of the delta peak that is the first term of Eq.~(\ref{eq:freecond}) in the noninteracting case.

\subsection{Infra-red approximation}
\begin{figure}[t]
\centering
\vskip 10pt
\includegraphics[width=70mm]{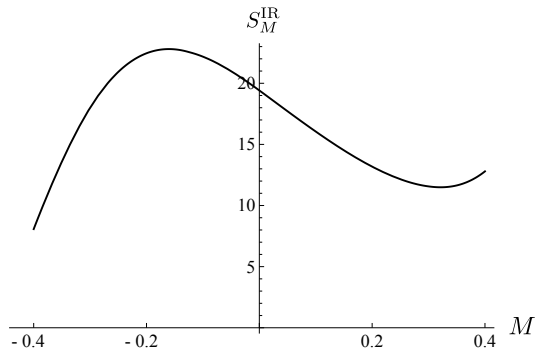}
\caption{The numerical value of the dimensionless integral $S^{\text{IR}}_M$ defined in the text as a function of $M$.}
\label{fig:SIR}
\end{figure}

In order to scrutinize the value of the conductivity in the dc limit $\hbar\omega/k_B T \ll 1$, we firstly introduce dimensionless variables $x=\hbar\omega \beta$ and $y=c\hbar|\mathbf{k}|\beta$. As can be inferred from Eq.~(\ref{eq:xi}), the eigenvalues $\xi_{\pm}$ scale as $ \lim_{r\rightarrow \infty} (r/\ell^2)^{2M} \xi_{\pm}(\omega,|\mathbf{k}|) =(\hbar c\beta)^{-2M} s_M(x,\pm y)$, with $s_M$ an $M$-dependent dimensionless function. Next, we observe that in the far infra-red, the Green's function is dominated by the self-energy term. We can approximate it by neglecting the kinetic part, so that the Green's function inherits the $1/T^{2M}$ temperature scaling from the self-energy. This approximation is valid for $\hbar \omega \ll k_B T \ll \hbar \omega_{\text{c}}$.
Proceeding to compute the conductivity as before, we now obtain in this limit the following expression for the total dc conductivity:
\begin{equation}
\sigma^{\text{IR}}=\lim_{\hbar\omega\beta\rightarrow 0}\text{Re}\,\sigma_{xx}(\mathbf{0},\omega) \simeq \frac{e^2 \omega_{\text{c}}}{12\pi\hbar c} \left(\frac{k_B T}{\hbar \omega_c}\right)^{3-4M} S_M^{\text{IR}},\nonumber
\end{equation}
where $S_M^{\text{IR}}$ is a dimensionless integral that in the limit $\hbar\omega \beta\rightarrow 0$ is given by
\begin{align}
&S_M^{\text{IR}}=\left[\frac{\Gamma(\frac{1}{2}-M)}{4^{M}\Gamma(\frac{1}{2}+M)}\right]^2\int_0^{\infty}dy \int_{-\infty}^{\infty} dx \frac{y^2  }{\cosh^2(x/2)}\nonumber \\
&\times\Big(4\mathcal{A}^{\text{IR}}_+(x,y) \mathcal{A}^{\text{IR}}_-(x,y) + \left[\mathcal{A}^{\text{IR}}_+(x,y)\right]^2+ \left[\mathcal{A}^{\text{IR}}_-(x,y)\right]^2  \Big) .\nonumber
\end{align}
Here the components of the spectral-weight function in the far infra-red are given in terms of the functions $s_M$ by
$\mathcal{A}^{\text{IR}}_{\pm}(x,y)={\rm Im} \left[-1/\pi s_M(x+i0,\pm y)\right]$. The numerical value of this integral is shown in Fig.~\ref{fig:SIR} for various $M$.
The conductivity thus indeed tends to a constant in the far infra-red, and scales as $\sigma^{\text{IR}}\propto T^{3-4M}$. This is true for both contributions, i.e., both the DC value of the interband part and the height of the Drude peak scale as $T^{3-4M}$.
The above scaling argument also suggests that the width $\Gamma$ of the Drude peak scales with temperature in the same way as the self-energy, i.e., $\Gamma(T) \propto T^{2M}$, so that the Drude peak behaves as $ T^{3-2M}\Gamma/(\Gamma^2 + \omega^2)$. Our numerics are indeed consistent with this scaling.

\section{Discussion}
In this work, we have obtained a model for strongly coupled Dirac semimetals with holographic interaction effects. Using this model, we have computed the fermionic contribution to the electrical conductivity. We have shown that for small frequency, it inherits the scaling of the critical order-parameter field responsible for the interactions between the Dirac fermions.
The reason that the electrical conductivity is finite at the charge neutrality point, is that an external electric field couples to particle and hole currents with opposite sign. This fact, combined with particle-hole symmetry, makes sure that an electric field does not couple to the center-of-mass motion of the system. In addition, the interactions lead to a drag force that relaxes the charge current in finite time, making the conductivity finite. This is in contrast to the thermal conductivity. Indeed, because of the linear dispersion, the heat current is directly proportional to the center-of-mass momentum. Therefore interactions do not relax the heat current driven by a thermal gradient, i.e., the thermal conductivity of a Dirac semimetal is infinite \cite{Muller09}.

A closely related consequence of particle-hole symmetry is the fact that vertex corrections are not crucial for obtaining the qualitative behavior of the conductivity. The charge transport relaxation time is finite, and in the absence of vertex corrections, it is approximated by the single-particle lifetime. This decreases the final result for the conductivity by a multiplicative numerical factor. The latter represents an angular effect, taking into account that not all scattering events contributing to a finite lifetime, contribute to current relaxation equally effectively \cite{Mahan00}. In fact, we expect this angular correction to be small in the strongly coupled case, where there is no preference for forward scattering, as opposed to weakly coupled systems where small-angle scattering is dominant. Therefore, our results in the $GG$ approximation can be interpreted as a lower bound on the exact result, possibly with different numerical coefficients in the scaling laws, but with the same universal features. In particular, the facts that the UV limit yields the free result, that the IR limit is a scaling law with power $3-4M$, that there is a Drude peak, and that one or two crossovers are present in between these regimes, are all universal features that are not sensitive to vertex corrections.
Computation of the vertex corrections requires further study. In particular, we need information on the dependence of the self-energy on the dressed Green's function, which is difficult to obtain in the present model and this is beyond the scope of this work.

Holographic methods are usually entirely universal, in the sense that they can be used to describe classes of possible condensed-matter systems. The advantage is that our model is very general. However, it remains a challenge to predict the behavior of specific realistic systems. We expect the obtained power-law behavior of the conductivity to be applicable to a wide range of condensed-matter systems in a universality class determined by $M$, but its coefficient to depend on material specifics. For instance, in our model, the latter depends on the Fermi velocity $c$, which must be replaced by its appropriate experimental value. The Fermi velocities are of the order $10^{5}$ and $10^{6}$ m/s in Na$_3$Bi and Cd$_3$As$_2$, respectively \cite{Chen14,Neupane13}.

Hopefully, also future experiments will reveal the behavior of Dirac semimetals coupled to fluctuating collective modes. Comparing holographic models with experimental data will then lead to a better understanding of the relation between string-theoretic methods and realistic condensed-matter systems.

\acknowledgments{}
We thank Watse Sybesma, Lars Fritz, Simonas Grubinskas, and Umut G\"ursoy for useful and stimulating discussions. This work was supported by the Stichting voor Fundamenteel Onderzoek der Materie (FOM), the Netherlands Organization for Scientific Research (NWO) and the Delta-Institute for Theoretical Physics (D-ITP).

\bibliography{dirac}

\end{document}